\documentclass[11pt]{article}

\usepackage{indentfirst}
\usepackage{graphicx}
\usepackage{longtable}
\usepackage{supertabular}
\usepackage{amsmath}
\usepackage{amssymb}
\usepackage{color}
\usepackage{amstext}
\usepackage{cases}
\usepackage{float}
\usepackage{threeparttable}
\usepackage{booktabs}
\usepackage[margin=.8in]{geometry}
\usepackage{setspace}
\usepackage{multirow}
\usepackage{array}
\usepackage{mathrsfs}
\usepackage{enumitem}
\setlist[enumerate]{listparindent=\parindent}

\begin{document}

\title{{\Large\bf {Conditions for Social Preference Transitivity When Cycle Involved and A $\widehat{O}\mbox{-}\widehat{I}$ Framework}} }

\author{Fujun Hou \thanks{Email: houfj@bit.edu.cn.} \\
School of Management and Economics\\
Beijing Institute of Technology\\
Beijing, China, 100081}
\date{}
\maketitle

\begin{abstract}
  We present some conditions for social preference transitivity under the majority rule when the individual preferences include cycles. First, our concern is with the restriction on the preference orderings of individuals except those (called cycle members) whose preferences constitute the cycles, but the considered transitivity is, of course, of the society as a whole. In our discussion, the individual preferences are assumed concerned and the cycle members' preferences are assumed as strict orderings. Particularly, for an alternative triple when one cycle is involved and the society is sufficient large (at least 5 individuals in the society), we present a sufficient condition for social transitivity; when two antagonistic cycles are involved and the society has at least 9 individuals, necessary and sufficient conditions are presented which are merely restricted on the preferences of those individuals except the cycle members. Based on the work due to Slutsky (1977) and Gaertner \& Heinecke (1978), we then outline a conceptual $\widehat{O}\mbox{-}\widehat{I}$ framework of social transitivity in an axiomatic manner. Connections between some already identified conditions and the $\widehat{O}\mbox{-}\widehat{I}$ framework is examined.

\end{abstract}

\setlength{\unitlength}{1mm}

\section{Introduction}

 In social choice theory, if the individuals provide their preferences without any restriction, then, social preference inconsistency may arise. A famous example is the voting paradox. This is one of the results of Arrow's Impossibility Theorem (Arrow, 1950,1951). To avoid the voting paradox, people have identified conditions imposed on the society's preference profile permitting any number of the individuals to select their preferences if only their preferences fall into the required domain (Arrow,1951; Black,1958; Inada,1964,1969; Ward, 1965; Sen, 1966; Sen \& Pattanaik, 1969; Duggan,2016; and many others).

 The collection of already identified conditions is proved complete in literature such as, particularly, the condition list of Inada (1969) for transitive social ordering, and the conditions of value restriction (VR), extremal restriction (ER) and limited agreement (LA) of Sen \& Pattanaik (1969) for both transitive social ordering and non-empty social choice set.

 It is legitimate to enquire whether the condition list of Inada (1969) and the conditions of VR, ER and LA considered by Sen \& Pattanaik (1969) cover all the cases where the majority rule always yield transitive orderings. The answer is no. There are indeed such examples that neither condition in the list of Inada nor condition of VR, ER or LA is satisfied, but still able to produce transitive social orderings under the majority rule. For one example, Saposnik (1975) presented the condition of cycle balance (CB) guaranteeing social preference transitivity when the individual preference set contains merely opposite cycles (namely, clockwise cycles and counterclockwise cycles). For another, Gaertner \& Heinecke (1978) presented the condition of cyclically mixed preferences (CM). The CM is restricted on a reduced preference structure which ultimately contains, regardless of distribution number, strict orderings (if any) all belonging to the same $U$-cycle ($U_i$-cycles of Gaertner \& Heinecke, $i=1,2$, correspond to Saposnik's clockwise cycle and counterclockwise cycle, respectively), and at most of two types of weak orderings belonging to different $V$-cycles (The 6 weak orderings (except the unconcerned) over an alternative triple are separated into two cycles by Gaertner \& Heinecke).

 In the situation where the individual preference set contains cycles, this paper first investigate the restriction on the preference orderings of individuals except those (called cycle members) whose preferences constitute the cycles but the overall social preference transitivity is still anticipated under the majority decision rule. And then we make, in an axiomatic manner, a try to outline a $\widehat{O}\mbox{-}\widehat{I}$ framework for social transitivity.

\section{Preliminary}

In this section, we introduce some preliminaries relevant to our discussion.

\subsection{Assumption, notation and definition}

The individual preferences are assumed to be weak orderings over a finite set of alternatives. We further assume that the number of alternatives is more than 2 and that the individuals number is not less than 5 (to be explained in subsection 4.1).

As usual, for two alternatives $x$ and $y$, individual $i$'s preference "$x$ is at least as good as $y$" is denoted by $x R_i y$, "$x$ is preferred to $y$" is denoted by $x P_i y$, and "$x$ is indifferent to $y$" is denoted by $x I_i y$. Similarly, $R_C$, $P_C$ and $I_C$ are used to denote the collective preference of the society. A \textit{concerned individual} is a person who is not indifferent between all the alternatives (Sen, 1966). Accordingly, \textit{unconcerned preferences} and \textit{concerned preferences} can thus be distinguished.

Let $N(xRy)$, $N(xPy)$ and $N(xIy)$ denote the numbers of individuals regarding $x R_i y$, $x P_i y$ and $x I_i y$, respectively. Sometimes, when the numbers are of the preferences of individuals in a given set $S$, we use such notations as $N_S(xRy)$, $N_S(xPy)$ and $N_S(xIy)$.

The \textit{simple majority decision rule} means:
\begin{itemize}
\item [$\diamond$] $xRy$ if and only if $N(xRy)\geq N(yRx)$,
\item [$\diamond$] $xPy$ if and only if $N(xRy)> N(yRx)$,
\item [$\diamond$] $xIy$ if and only if $N(xRy)= N(yRx)$.
\end{itemize}

For a triple of alternatives $x,y,z$, the \textit{social transitivity} means: If $xRy$ and $yRz$ then $xRz$. Particularly,
\begin{itemize}
\item If $xPy$ and $yPz$ then $xPz$,
\item If $xPy$ and $yIz$ then $xPz$,
\item If $xIy$ and $yPz$ then $xPz$,
\item If $xIy$ and $yIz$ then $xIz$.
\end{itemize}

A \textit{strict cycle} is defined in this paper by three strict orderings,
$$xPyPz, yPzPx, zPxPy.$$
The individuals who hold the preferences in a cycle are called \textit{cycle members}, and their preferences are called \textit{cyclic preferences}. Moreover, we use $Y$ to denote a set of cycle members.

An \textit{antagonistic preference pair} is defined as $\{xPyPz,zPyPx\}$ (Inada, 1964). Similarly, we define \textit{antagonistic weak preference pair} by $\{xPyIz,zIyPx\}$. We further define \textit{antagonistic cycles} by
$$\begin{array}{ll}
\left\{
 \begin{array}{ll}
    xPyPz\\
    yPzPx\\
    zPxPy
 \end{array} \right.\&
 \left\{
 \begin{array}{ll}
    zPyPx\\
    xPzPy\\
    yPxPz
 \end{array}
\right..
\end{array}$$

We note that, Saposnik (1975) distinguishes the above two cycles by clockwise cycle and counterclockwise cycle, and Gaertner \& Heinecke (1978) denotes them by $U_1$-cycle and $U_2$-cycle, respectively.

As well known, consistent decision and rational choice are fundamental in social choice theory. For a set of individual preferences (a preference profile), however, the majority decision rule may not yield a transitive social preference ordering or a non-empty social choice set (Arrow,1950,1951; Sen \& Pattanaik,1969). In the following two subsections, we will list some conditions related to social preference ordering.

\subsection{Conditions listed by Inada}

Inada (1969) listed the conditions which he proved that the collection is complete, that is, the conditions he listed cover all possible conditions for social transitivity under the majority rule. For the convenience of readers, we reproduce Inada's description as follows.

\textbf{Inada's Theorem (Inada,1969):} For the simple majority decision rule to yield a transitive social preference, the individual preferences over any alternative triple $(x,y,z)$ should satisfy at least one of the following patterns:
\begin{itemize}
\item [(i)] dichotomous preferences (number of voters is free): $xPyPz$ do not appear as possible individual preferences;
\item [(ii)] echoic preferences (number of voters is free): if an individual has preference $xPyPz$, then $zPx$ is not allowed for anyone else;
\item [(iii)] antagonistic preferences (number of voters is free): if an individual has preference $xPyPz$, then the preference of anyone else should be $zPyPx$ or $xIz$;
\item [(iv)] value-restricted preferences (number of voters is odd): there is one alternative and one value ("best," "worst," or "medium") such that the alternative never has that value in any concerned individual's preference;
    \begin{itemize}
    \item [(a)] single-peaked preferences,
    \item [(b)] single-caved preferences,
    \item [(c)] two-group-separated preferences,
    \end{itemize}
\item [(v)] taboo preferences (number of voters is odd): $xIyIz$ does not appear, and $x$ is "best" or $y$ is "worst" for any individual orderings.
\end{itemize}

As sufficiency conditions, we remark that, the single-peaked restriction was first proposed by Black (1948, 1958) and later formally by Arrow (1951), the value restriction was proposed by Sen (1966), and the other restrictions in the above list were proposed by Inada (1964, 1969).

\subsection{Conditions of VR, ER, and LA}

Sen and Pattanaik (1969) considered the following conditions:
\begin{itemize}
    \item [(i)] value restriction (VR): as aforementioned;
    \item [(ii)] extremal restriction (ER): for an ordered triple $(x,y,z)$ of alternatives, $[\exists i: x P_i y ~\&~ y P_i z]\rightarrow [\forall j: z P_j x\rightarrow z P_j y ~\&~ y P_j x]$;
    \item [(iii)] limited agreement (LA): in a triple of distinct alternatives there is an ordered pair $(x,y)$ such that $\forall i: xR_iy$;
\end{itemize}
where ER combines three patterns, i.e., dichotomous preferences, echoic preferences and antagonistic preferences; and LA is a weaker version of the pattern of taboo preferences.

Sen and Pattanaik (1969) applied VR, ER and LA to several theorems regarding necessary and/or sufficient conditions for social preference transitivity. We reproduce one of them with the original reference number remained.

\textbf{Theorem XI (Sen and Pattanaik (1969))} \textit{The necessary and sufficient condition for a set of individual orderings to be in the domain of majority-decision-SWF is that every triple of alternatives must satisfy extremal restriction.}

The terms of SWF is the abbreviation for "social welfare function".

\subsection{Condition of CB}

The condition of cycle balance (CB) is proposed for a situation where the individual preference set simply includes pairs of clockwise cycle and counterclockwise cycle. Saposnik (1975) proves that, the majority decision rule always yields a transitive social ordering when individual preferences satisfy the CB. It is quite easy to understand this assertion since each pair of antagonistic groups reduces to a unconcerned preference.

\subsection{Condition of CM}

The condition of cyclically mixed preferences (CM) is a restriction imposed on a reduced preference structure of the original preference profile (Gaertner \& Heinecke, 1978). Here we briefly outline the chief components of CM. We start with the separation of 12 possible orderings (disregarding the unconcerned one) over an alternative triple $(x,y,x)$:
$$\begin{array}{ll}
 U_1\text{-cycle}: & \{x P y P z, y P z P x, z P x P y\},\\
 U_2\text{-cycle}: & \{z P y P x, x P z P y, y P x P z\},\\
 V_1\text{-cycle}: & \{x P y I z, y P z I x, z P x I y\},\\
 V_2\text{-cycle}: & \{z I y P x, x I z P y, y I x P zy\}.
\end{array}
$$

The following observations are of use in the establishment of CM:
\begin{itemize}
\item [(O1)] The joint occurrence of $U_1$ and $U_2$ is equivalent to an unconcerned preference. In addition, two elements corresponding to the upper and lower positions with respect to $U_1$ and $U_2$ together imply an unconcerned preference. In other words, the antagonistic (opposite) relationship holds pairwise with respect to $U_1$ and $U_2$.
\item [(O2)] Each $V_i$ implies, respectively, an unconcerned preference. In addition, two elements corresponding to the upper and lower positions with respect to $V_1$ and $V_2$ together imply an unconcerned preference. In other words, the antagonistic (opposite) relationship holds pairwise with respect to $V_1$ and $V_2$. Therefore, a group of two distinct elements in $V_i$ can be replaced by the antagonistic (opposite) ordering (in $V_j$, $j\neq i$) of the third element in $V_i$.
\item [(O3)] Consider two distinct weak orderings in same $V$-cycle. Based on O2, it is clear that these two distinct weak orderings are equivalent to the weak ordering in the other $V$-cycle which is none of the two antagonistic weak orderings of the two weak orderings under consideration.
\item [(O4)] Two strict orderings (not in antagonistic relationship) from different $U$-cycles can be replaced by 'a weak ordering from a  $V$-cycle but counted twice'. For instance, $\{xP_1yP_1z,xP_2zP_2y\}$ can be replaced by $\{zI_3yP_3x,zI_4yP_4x\}$ under the majority rule.
\end{itemize}

To obtain the ultimate reduced structure, a sequence of 4 procedures, which are based on the above 4 observations, is implemented in an iteratively way until the profile under reduction does not change any longer. The original preference structure is equivalent to a unique steady reduced structure. For an alternative triple $(x,y,z)$, Gaertner \& Heinecke (1978) proves,  that the steady reduced structure reaches one and only one of 12 different standard forms denoted by $B_{k,l}$, where $k$ ($k=0,1,2,3$) is the number of strict orderings from a same $U$-cycle, while $l$ ($l=0,1,2$) is the number of weak orderings from different $V$-cycles (if $l=2$, the two weak orderings are, of course, not an antagonistic pair). Then Gaertner \& Heinecke (1978) establishes the CM condition for social transitivity with respect to 8 cases (except $B_{2,0},B_{3,0},B_{3,1},B_{3,2}$). For more detail, one may refer to Gaertner \& Heinecke (1978).

\section{An example and observation}

\subsection{Example}

We consider the following example.

\textbf{Example 1} Suppose that 5 voters provide their preference orderings over an alternative triple $(x,y,z)$ as follows.
$$\begin{array}{ll}
 \text{Voter 1}: & x I_1 y P_1 z,\\
 \text{Voter 2}: & x P_2 y I_2 z,\\
 \text{Voter 3}: & x P_3 y P_3 z,\\
 \text{Voter 4}: & y P_4 z P_4 x,\\
 \text{Voter 5}: & z P_5 x P_5 y.
\end{array}
$$

Applying the majority decision rule to the aggregation of the individual preferences, we obtain a transitive social ordering $x P_C y P_C z$ since
$$\begin{array}{ll}
 N(xPy)=3, & N(yPx)=1,\\
 N(yPz)=3, & N(zPy)=1,\\
 N(xPz)=3, & N(zPx)=2.
\end{array}$$

\subsection{Observation}

Now we check whether the conditions listed in Section 2 are fulfilled by the individual preferences in Example 1.
\begin{itemize}
\item [(1)] Whether the conditions listed by Inada (1969) are satisfied by the profile in Example 1?
\begin{itemize}
\item [(i)] dichotomous preferences (number of voters is free): VIOLATED, since $x P_3 y P_3 z$ appears as the voter 3's preference;
\item [(ii)] echoic preferences (number of voters is free): VIOLATED, since we have both $x P_3 y P_3 z$ and $z P_4 x$;
\item [(iii)] antagonistic preferences (number of voters is free): VIOLATED, since we have both $x P_3 y P_3 z$ and $z P_5 x P_5 y$;
\item [(iv)] value-restricted preferences (number of voters is odd): VIOLATED, since any of the three alternatives can be ranked at any position from $\{1,2,3\}$ by the last three voters;
    \begin{itemize}
    \item [(a)] single-peaked preferences,
    \item [(b)] single-caved preferences,
    \item [(c)] two-group-separated preferences,
    \end{itemize}
\item [(v)] taboo preferences (number of voters is odd): VIOLATED, since we have $x P_3 y P_3 z$, $y P_4 z P_4 x$ and $z P_5 x P_5 y$.
\end{itemize}
\item [(2)] Whether the conditions considered by Sen and Pattanaik (1969) are satisfied by the profile in Example 1?
\begin{itemize}
    \item [(i)] value restriction (VR): VIOLATED, as shown above;
    \item [(ii)] extremal restriction (ER): VIOLATED, because for one instance, we have $x P_3 y P_3 z$ and $z P_4 x$ but we do NOT have $z P_4 y P_4 x$.
    \item [(iii)] limited agreement (LA): VIOLATED, since we have $x P_3 y P_3 z$, $y P_4 z P_4 x$ and $z P_5 x P_5 y$.
\end{itemize}
\item [(3)] The profile in Example 1 does not satisfy the condition CB of Saposnik (1975) since there is only one cycle in the profile.
\item [(4)] The profile in Example 1 does not satisfy the condition CM of Gaertner \& Heinecke (1978) since the profile belongs to the stand form of $B_{3,2}$, which is not cyclically mixed.
\end{itemize}

One can observe that none of the conditions listed in Section 2 is fulfilled by the individual preferences in Example 1.

Is Example 1 a counter example to those well-known conditions? The answer is no.
Sen \& Pattanaik put it very clear (Sen \& Pattanaik, 1969), that, the conditions of VR, ER and LA are about the preference pattern rather than on the individual number, that is, any number of individuals can select their preferences if only their preferences satisfy the conditions then the majority rule will yield a transitive social ordering.

In the next section, we will present some conditions for social preference transitivity under the majority rule when the individual preferences include cycles.

\section{Some conditions when cycle involved}

\subsection{A special case when one cycle included}

We consider a special case where a single cycle is included in the individual preferences. We first give a lemma.

\textbf{Lemma 1} For a triple of distinct alternatives $(x,y,z)$ and a cycle over this triple, to destroy the cycle we need at least 2 additional orderings over that triple, that is, we need to add at least 2 orderings so that we may obtain a social transitivity over that triple when we consider the cycle members's preferences together with the to be added preferences under the majority rule.

\textbf{Proof} Without loss of generality, we consider the following preferences over $(x,y,z)$:
 $$\begin{array}{ll}
 \text{Individual 1}: & x P_1 y P_1 z,\\
 \text{Individual 2}: & y P_2 z P_2 x,\\
 \text{Individual 3}: & z P_3 x P_3 y,
 \end{array}
 $$
 which constitute a cycle due to $N(xPy)=2$, $N(yPz)=2$ and $N(zPx)=2$. To destroy the cycle we need at least 2 additional $xPz$ to eliminate the influence of $N(zPx)=2$ (note that when two additional orderings are introduced, the group under consideration will include 5 individuals rather than 3 any longer). Since one ordering can provide at most one $xPz$, hence at least 2 additional orderings are needed to reverse $zPx$. Therefore, Lemma 1 is proved.

 One can see from Lemma 1 that, in order to avoid a possible social preference cycle, a decision committee ought to be composed of at least 5 members. Lemma 1 explains why we assume that the society includes at least 5 individuals in subsection 2.1.

 Now we consider, under what condition can an alternative triple arrive at a transitive social ordering under the majority rule when cyclic preferences are involved. We have the following theorem.

 \textbf{Theorem 1} Suppose that there is a single cycle over triple $(x,y,z)$, and $xPyPz$ is in the cycle. Denote by $S=\{1,2,\cdots,n\}$ with $n\geq 5$ the set of concerned individuals over triple $(x,y,z)$, and denote by $Y$ the cycle member set. If the following condition is fulfilled
 $$\forall j: [j\in S\setminus Y] \rightarrow [xR_j y ~\&~ yR_j z ~\&~ xP_j z],\eqno (1)$$
the social preference over the triple $(x,y,z)$ is transitive under the majority decision rule.

\textbf{Proof} Without loss of generality, we assume the following cyclic preferences over $(x,y,z)$:
 $$\begin{array}{ll}
 \text{Individual 1}: & x P_1 y P_1 z,\\
 \text{Individual 2}: & y P_2 z P_2 x,\\
 \text{Individual 3}: & z P_3 x P_3 y.
 \end{array}
 $$
 Under this assumption, the set of cycle members is $Y=\{1,2,3\}$.

 First, we confine our attention only on the above cyclic preferences, we have
 $$
 \left\{
 \begin{array}{ll}
 N_Y(xPy)=2, N_Y(yPx)=1,\\
 N_Y(yPz)=2, N_Y(zPy)=1,\\
 N_Y(xPz)=1, N_Y(zPx)=2.
 \end{array}
 \right.
 $$

Second, we extend our attention to the individuals set $\{1,2,3,4,5\}$. Because we assume that the individual preferences are weak orderings and concerned over the triple $(x,y,z)$, thus when condition (1) is satisfied, we have
$$
 \left\{
 \begin{array}{ll}
 N_{\{1,2,3,4,5\}}(xPy)> N_{\{1,2,3,4,5\}}(yPx),\\
 N_{\{1,2,3,4,5\}}(yPz)> N_{\{1,2,3,4,5\}}(zPy),\\
 N_{\{1,2,3,4,5\}}(xPz)= 3 ~\&~N_{\{1,2,3,4,5\}}(zPx)=2.
 \end{array}
 \right.
 $$

Finally, we consider the contribution of each individual beyond $\{1,2,3,4,5\}$. Clearly, when condition (1) is satisfied, each individual beyond $\{1,2,3,4,5\}$ contributes to $N(xPy)$ at least as great as to $N(yPx)$, contributes to $N(yPz)$ at least as great as to $N(zPy)$, and contributes more to $N(xPz)$ than to $N(zPx)$. Hence we have
$$
 \left\{
 \begin{array}{ll}
 N_S(xPy)> N_S(yPx),\\
 N_S(yPz)> N_S(zPy),\\
 N_S(xPz)> N_S(zPx),
 \end{array}
 \right.
 $$
 which indicates a transitive social ordering $xP_SyP_Sz$. We complete the proof.

\subsection{A special case when two cycles included}

We consider a special case where two antagonistic cycles are included in the individual preferences. We have the following corollary.

\textbf{Corollary 1} (from Theorem XI of Sen \& Pattanaik (1969)) Suppose that in the individual preference set there two antagonistic cycles over triple $(x,y,z)$, and their cycle member sets are denoted by $Y_1$ and $Y_2$. Denote by $S=\{1,2,\cdots,n\}$ with $n\geq 9$ the set of concerned individuals over triple $(x,y,z)$. The necessary and sufficient condition for the social preference over the triple $(x,y,z)$ to be transitive under the majority decision rule is that, the preferences of individuals except those from $Y_1\cup Y_2$ satisfy ER.

\textbf{Proof} The theorem is quite intuitive, since the antagonistic cycles over $(x,y,z)$ together imply $xIyIz$ holding locally with respect to the preferences of individuals from $Y_1\cup Y_2$. From the Theorem XI of Sen \& Pattanaik (1969), on the other hand, we know that the ER is the necessary and sufficient condition for the preferences of individuals from $S\setminus (Y_1\cup Y_2)$ to be of local social preference transitivity. And thus the social preference transitivity over $(x,y,z)$ can be also obtained. We complete the proof.

Following Inada's theorem (1969), we can obtain another corollary.

\textbf{Corollary 2} (from Inada's theorem (1969)) Suppose that in the individual preference set there two antagonistic cycles over triple $(x,y,z)$, and their cycle member sets are denoted by $Y_1$ and $Y_2$. Denote by $S=\{1,2,\cdots,n\}$ with $n\geq 9$ the set of individuals over triple $(x,y,z)$. The necessary and sufficient condition for the social preference over the triple $(x,y,z)$ to be transitive under the majority decision rule is that, at least one of the Inada condition list is satisfied by the preferences of individuals except those from $Y_1\cup Y_2$.

\section{A $\widehat{O}\mbox{-}\widehat{I}$ framework of social transitivity}

In this section, we outline a conceptual framework of social transitivity. This section uses some notations of its own most of which are set relevant and in a majority sense.
\begin{itemize}
    \item [-] $S=\{1,2,\ldots,n\}$ stands for a society with $1<n<+\infty $;
    \item [-] $G$ stands for a group where $G\subseteq S$;
    \item [-] $A$ stands for an alternative set with $1<|A|<+\infty $; and suppose $X\subseteq A$
    \item [-] Let $\Gamma(A)$ denote the set of possible orderings over $A$. $\Gamma(X)$ has the similar meaning but with respect to $X$;
    \item [-] Define $\Pi(S,A)$ as a preference profile of $S$ over $A$ such that $\Pi(S,A)\subseteq S\times\Gamma(A)$, $|\Pi(S,A)|=|S|$, and, if $\{[(i,a)\in \Pi(S,A)]\& [(j,b)\in \Pi(S,A)]\}$ then $i\in S$, $j\in S$ and $i\neq j$; that is, $\Pi(S,A)$ includes $|S|$ elements, and the first entries of its elements constitute a permutation of the elements in $S$. $\Pi(G,A)$, $\Pi(G,X)$  and $\Pi(S,X)$ are defined in a similar way;
    \item [-] Let $\Omega(S,A)$ denote the set of all possible profiles of $S$ over $A$;
    \item [-] Denote the unconcerned preference by $UCP$. In addition, we do not distinguish a $UCP$ over $A$ and a $UCP$ over $X$;
    \item [-] $M(\bullet)$ represents the majority aggregation operation on a preference profile such that a collective preference (not necessary an ordering) can be obtained;
    \item [-] $\widehat{I_{G,A}}$ is called an indifferent set of $G$ over $A$ such that $\widehat{I_{G,A}}\in \Omega(G,A)$ and $M(\widehat{I_{G,A}})=UCP$;
    \item [-] $\widehat{O_{G,A}}$ is called an ordinal set of $G$ over $A$ such that $\widehat{O_{G,A}}\in \Omega(G,A)$ and $M(\widehat{O_{G,A}})\in \Gamma(A)$, that is, $\widehat{O_{G,A}}$ yields a transitive collective ordering under the majority rule;
    \item [-] The symbol '+' represents the union operator of sets;
    \item [-] $D_{G_1,A}$ and $D_{G_2,A}$ are called opposite sets such that
        \begin{itemize}
        \item [(i)] $D_{G_1,A}\in \Omega(G_1,A)$, $D_{G_2,A}\in \Omega(G_2,A)$;
        \item [(iii)] $G_1 \cap G_2=\emptyset$ and $|G_1|+|G_2|\leq n$;
        \item [(iv)] $M(D_{G_1,A}+ D_{G_2,A})=UCP$.
        \end{itemize}
\end{itemize}

We postulate four axioms each attached a conceptual interpretation next to it.
\begin{enumerate}
\item [] \textbf{Axiom 1 (Non-preference Axiom)} $M(\emptyset)=UCP$.

No preference implies no influence on the final decision outcome.
\item [] \textbf{Axiom 2 (Additivity Axiom)} $M(\widehat{O_{G_1,A}}+ \widehat{I_{G_2,A}})=M(\widehat{O_{G_1,A}})$, where $G_1\cap G_2=\emptyset$ and $|G_1|+|G_2|\leq n$.

Adding an indifferent set to an ordinal set does not influence the original outcome.
\item [] \textbf{Axiom 3 (Subtractivity Axiom)} $M(\widehat{O_{G_1,A}}- \widehat{I_{G_2,A}})=M(\widehat{O_{G_1,A}})$, where $G_2\subseteq G_1$.

    Subtracting an indifferent set from an ordinal set does not influence the original outcome.
\item [] \textbf{Axiom 4 (Substitutability Axiom)} If both $D_{G_2,A}$ and $D_{G_3,A}$ are the opposite sets of $D_{G_1,A}$, then $M(D_{G_2,A})=M(D_{G_3,A})$.

    This item can be clarified by an example. Consider $\{(1, x P y I z), (2, y P z I x), (3, z P x I y)\}$ and $\{(1,x P y I z), (4,z I y P x)\}$ both being indifferent sets over $X=\{x,y,z\}$. The opposite sets of $\{(1,x P y I z)\}$ (denoted by $D_{\{1\},X}$) are $\{(2,y P z I x), (3,z P x I y)\}$ (denoted by $D_{\{2,3\},X}$) with respect to the former set and $\{(4,z I y P x)\}$ (denoted by $D_{\{4\},X}$) with respect to the later set. Clearly, under the majority rule we know that $\{(2,y P z I x), (3,z P x I y)\}$ and $\{(4,z I y P x)\}$ are equivalent, that is, $M( D_{\{2,3\},X})=M(D_{\{4\},X})$.
\end{enumerate}

The collection of the above axioms indicates a '$\widehat{O}\mbox{-}\widehat{I}$' framework, denoted $\langle [\Omega+\emptyset],\widehat{O}, \widehat{I}, M, D, UCP\rangle$, of social transitivity. It combines some statements of Slutsky (1977) and Gaertner \& Heinecke (1978) in an axiomatic manner with our concern particularly with the individual preference sets that yield transitive orderings under the majority rule. Consequently, many already identified conditions (Corollary 1 and Corollary 2 included) for social transitivity can be found in somewhat relationship with the $\widehat{O}\mbox{-}\widehat{I}$ framework:
\begin{itemize}
\item [$\diamond$] The VR embodies Axiom 3, since it disregards unconcerned individuals.
\item [$\diamond$] The antagonistic preferences condition embodies Axiom 2, since a preference set of this type may include many $\widehat{I}$s.
\item [$\diamond$] The CB is of the type indicated by Axiom 2, since each pair of clockwise cycle and counterclockwise cycle corresponds to a $\widehat{I}$.
\item [$\diamond$] Corollary 1 and Corollary 2 are of the type indicated by Axiom 2.
\item [$\diamond$] All the axioms are evidenced by the CM in its reduction process.
\end{itemize}

Moreover, the $\widehat{O}\mbox{-}\widehat{I}$ framework $\langle [\Omega+\emptyset],\widehat{O}, \widehat{I}, M, D, UCP\rangle$ might indicate, conceptually, many transitivity conditions by
\begin{itemize}
\item [(1)] seeking a $D_{G_1,X}$, constructing a $D_{G_2,X}$ such that $M(D_{G_1,X}+ D_{G_2,X})=UCP$, where $G_1 \cap G_2=\emptyset$ and $|G_1|+|G_2|\leq n$; and
\item [(2)] attaching a $\widehat{O_{G_3,X}}$ to $D_{G_1,X} + D_{G_2,X}$ such that $|G_1|+|G_2|+|G_3|\leq n$. We then know that $M(D_{G_1,X}+ D_{G_2,X}+ \widehat{O_{G_3,X}})\in \Gamma(X)$, namely, $D_{G_1,X}+ D_{G_2,X}+ \widehat{O_{G_3,X}}$ yields a transitive social ordering over $X$ under the majority rule.
\end{itemize}
For instance, one can formulate such transitive conditions as '$B_{3,1}$ balanced', '$B_{3,2}$ balanced', etc., where $B_{3,1}$ and $B_{3,2}$ belong to standard forms of Gaertner \& Heinecke (1978). By the way, the CB of Saposnik (1975) belongs to a '$B_{3,0}$ balanced'.

We remark that $\langle [\Omega+\emptyset],\widehat{O}, \widehat{I}, M, D, UCP\rangle$ is merely a conceptual framework rather than an algebra structure in mathematical sense.

\section{Concluding remarks}

In this paper, we presented some conditions for the individual preferences to yield a transitive social ordering when a single cycle or two antagonistic cycles are included. The conditions were established over alternative triples. In practice, the proposed conditions might be used to test social preference transitivity along with those already identified conditions. Then we made an attempt, based on the work due to Slutsky (1977) and Gaertner \& Heinecke (1978), to outline a conceptual $\widehat{O}\mbox{-}\widehat{I}$ framework for social transitivity, whose axioms were found effective for some existent conditions, and prospective for formulating new transitivity conditions.

\vspace{0.3cm}


\end{document}